\documentclass[epj-spec]{svjour}
\usepackage[dvips]{graphicx}
\usepackage{amsmath}
\usepackage{amssymb,dsfont,epsfig}

\newcommand{\imu}{\dot{\imath}}

\begin{document}
\title{Adiabatic cavity QED with pairs of atoms}
\subtitle{Atomic entanglement and Quantum teleportation}
\author{C.~Lazarou\thanks{\email{cl90@sussex.ac.uk}} \and B.M.~Garraway}
\institute{Department of Physics and Astronomy,
 University of Sussex, Falmer, Brighton, BN1 9QH, United Kingdom}
\abstract{
   We study the dynamics of a pair of atoms,
   resonantly interacting with a single mode
   cavity, in the situation where the atoms enter
   the cavity with a time delay between them.
   Using time dependent coupling functions to
   represent the spatial profile of the mode, we
   considered the adiabatic limit of the
   system. Although the time evolution is mostly
   adiabatic, energy crossings play an important
   role in the system dynamics.  Following from
   this, entanglement, and a procedure for cavity
   state teleportation are considered.  We examine
   the behaviour of the system when we introduce
   decoherence, a finite detuning, and potential
   asymmetries in the coupling profiles of the
   atoms.
   }
\maketitle

\section{Introduction} \label{sec:Intro}

Entanglement between Quantum systems is important for the realisation of a
Quantum Computer \cite{Nielsen}. In recent years many authors have proposed
different methods, based on cavity QED systems, for entangling atoms. Some
of these proposals use single resonant interactions \cite{Zheng2005} or
strongly detuned cavities \cite{Zheng2000,Jane2002,You2003b}, or the
adiabatic sequential passage of atoms through cavities
\cite{Marr2003,Yong2007}.  Other schemes use decoherence-free spaces and
continuous monitoring of the cavity decay for generating entangled states
\cite{Plenio1999,Beige2000a,Beige2000b}. Another approach is to make use of
photon polarisation measurements for characterising the final atomic state
\cite{Shorencen2003,Chen2003,Duan2003}.

In a recent work \cite{Lazarou2007}, we considered an entangling system
consisting of a pair of two-level atoms resonantly interacting with a single
mode cavity. Taking into account the sequential motion of the atoms through
the cavity and the spatial profile of the mode, we utilised an atom-cavity
interaction with identical time dependent coupling functions for both atoms.
The main feature for this resonant system, when considering the adiabatic limit,
was the existence of an energy crossing at the
vicinity of a temporal degeneracy. Furthermore, one of the atoms entangles to
the cavity, and the system evolution for large photon numbers has a resemblance
to the Jaynes-Cummings model. Based on these features, \emph{fairly} robust
methods for entangling the atoms, for quantum state mapping and implementing a
SWAP and a C-NOT gate were proposed.

Here we examine the system dynamics in a more general approach by considering
the possibility of a finite detuning between a the atoms and the cavity mode,
and the potential of having asymmetries in the coupling profiles of the two
atoms with respect to each other. Furthermore, we also study the role of
decoherence and how this affects the dynamics, but also the fidelity of the
proposed applications. As long as the detuning is relatively small, the system
behaves in a similar way as before. The atom that enters the cavity second becomes
entangled with the field mode, whereas the first atom does not. Furthermore one can
still map this entanglement onto the pair of atoms. As long as the detuning is
small the fidelity is relatively high.

On the other hand, for increasing values of the detuning various non-adiabatic
effects related to avoided crossings in the adiabatic spectrum, give rise to a
tri-partite entanglement between the cavity and both atoms. For detuning much
larger than the coupling strength, the cavity is disentangled from the system
with the atoms being entangled to each other, while the cavity is
only virtually excited \cite{Yong2007}.

The differences between the coupling functions for the two atoms, give rise to
somewhat different dynamics. Although the second atom still entangles to the
cavity, there are two paths which depend on the initial state of the second
atom. The system is then characterised by two mixing angles which are
functions of the coupling, the interaction time, the photon number and the
asymmetry factor of the two coupling profiles.

For a leaky cavity, the effects on the fidelity of the system, and generally
the system dynamics, are suppressed as long as a high $Q$ cavity is being
used. For example micromaser cavities with $Q\sim10^{10}$ could be used to
experimentally realise the proposed model with negligible effects on the
system.

In the following section we introduce the model and the corresponding
Hamiltonian, discuss in brief the resonant limit and propose a teleportation
protocol for field state transfer between cavities. In section \ref{sec:3}, a
detailed analysis for the case of asymmetric atomic coupling profiles is
presented. Furthermore we discuss the off-resonant limit, and in section
\ref{sec:4} we discuss decoherence effects and how these affect the
system. Finally in section \ref{sec:5} we conclude by summarising our results.
\section{The atom-cavity model}\label{sec:2}
\subsection{The Hamiltonian} \label{sec:21}

\begin{figure}[!t]
  \begin{center}
    \includegraphics[width=6.0cm,height=6.0cm]{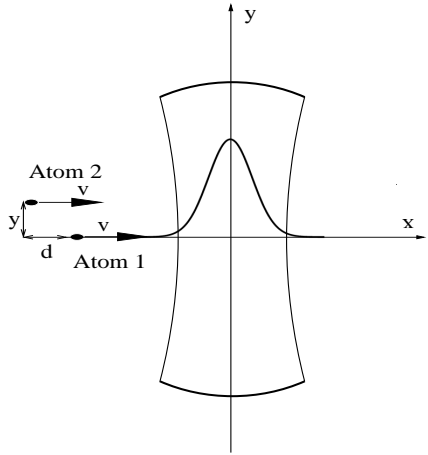}
      \caption{The experimental setup: Two identical two-level atoms enter a
    single mode cavity along two different trajectories. Both atoms are moving
    with the same velocity, but they enter and exit the cavity at different
    times. The field inside the cavity has a Gaussian profile along the $x$
    axis.}
      \label{fig:1}
  \end{center}
\end{figure}

In our model a pair of two-level atoms enters a single mode cavity at
different times, moving along the $x$ axis, atom 1, and
parallel to this, atom 2, figure \ref{fig:1}.
The field spatial profile has the form of a Gaussian function along the $x$
direction with additional spatial modulation along the $y$ direction
\begin{equation} \label{eq:1}
E(x,y)=E_{0}(y)\exp\big(-\frac{x^{2}}{4x_{0}^2}\big).
\end{equation}
Replacing the displacement operator with its classical counterpart,
$\hat{x}_j(t)\rightarrow v_jt$, we get the following coupling functions,
or profiles, for the
two atoms
\begin{equation} \label{eq:2}
\eta_{\scriptscriptstyle
1}(\tau)=g_1e^{-(\tau+\delta)^{2}},\qquad\eta_{\scriptscriptstyle
2}(\tau)=g_2e^{-(\tau-\delta)^{2}}.
\end{equation}
The dimensionless time $\tau$ and the parameter $\delta$ are defined
in terms of the time width $\sigma=x_{0}/v$
\begin{equation} \label{eq:3}
\tau=\frac{t}{2\sigma},\qquad\delta=\frac{\Delta t}{2\sigma}.
\end{equation}
The time interval $\Delta t$ is the time delay after which atom 2 enters the
cavity and $v$ the velocity of the atoms which is taken to be the same for
both atoms. We have already seen in Ref. \cite{Lazarou2007}, that the resonant
non-dissipative system is not too sensitive to the delay time $\delta$ and we
will rely on this in what follows.

Since atom 1 is moving along axis $x$ and atom 2 along a displaced trajectory,
figure \ref{fig:1}, and because the field strength depends on $y$ the coupling
for the second atom will be modified by a factor $\epsilon$ with respect to
$g_1=g_0$, $g_2=\epsilon g_1$. The parameter $\epsilon$ is the asymmetry
factor for the coupling profiles.

For the Eq. (\ref{eq:3}) to be valid the two atoms must be moving sufficiently
fast so that they don't get reflected by the cavity. This means that the
atomic velocity $v$ must be mush greater than the barrier set by the maximum
coupling strength $g_0$, $v\gg\sqrt{2\hbar g_0/m}$, where $m$ is the
atomic mass. Then, if this is the case the centre of mass motion is considered
to be classical and one can make the substitution $\hat{x}_j(t)\rightarrow vt$.

In the interaction picture, and within the rotating wave approximation, the
Hamiltonian reads $(\hbar=1)$
\begin{equation} \label{eq:4}
  H_{I}(t)=\Delta\left(\vert e_1e_2\rangle\langle e_1e_2\vert-\vert
  g_1g_2\rangle\langle
  g_1g_2\vert\right)+\sum_{j=1}^{2}\eta_{j}(t)\big(a^{\dagger}
  \sigma^{j}_{-}+a\sigma^{j}_{+}\big).
\end{equation}
The operators $a^{\dagger}$ and $a$ are the Bosonic
creation-annihilation operators  for the cavity mode, and
$\sigma^{j}_{-}$ and $\sigma^{j}_{+}$ are the lowering-raising
operators for atom $j$. The detuning between the atomic transition and the
mode frequency is $\Delta$. The ground state for atom $j$ is $\vert
g_j\rangle$ and the corresponding excited is $\vert e_j\rangle$.

\subsection{Resonant limit} \label{sec:22}
In a previous work \cite{Lazarou2007}, we considered the adiabatic limit for the
Hamiltonian (\ref{eq:4}) with $\Delta=0$, and emphasis in the case of equal
coupling amplitudes $g_1=g_2=g_0$, although some of the results are
somewhat more general and apply in the case of different couplings
$g_1\neq g_2$.

In the adiabatic limit one can diagonalise the Hamiltonian assuming that the
time is a parameter to obtain the time dependent adiabatic energies and the
corresponding state vectors \cite{Messiah}. The only relevant bare states for
the purposes of our analysis are those with the
same number of total excitations since the total number of excitations is a
constant of motion
\begin{equation} \label{eq:5}
\vert n,e_{1}e_{2}\rangle,\quad\vert n+1,g_{1}e_{2}\rangle,\quad\vert
n+1,e_{1}g_{1}\rangle,\quad\vert n+2,g_{1}g_{2}\rangle .
\end{equation}

After diagonalising the Hamiltonian (\ref{eq:4}), we get the following
adiabatic energies and the corresponding state vectors \cite{Lazarou2007,Mahmood1987}
\begin{subequations} \label{eq:6}
\begin{align}
E_{1,2}(\tau)&=\mp E_{-}(\tau),\qquad E_{3,4}=\mp E_{+}(\tau),
\label{eq:6a} \\ \nonumber \\
E_{\pm}(\tau)&=\sqrt{\frac{(3+2n)(\eta_{\scriptscriptstyle
1}^{2}(\tau)+\eta_{\scriptscriptstyle 2}^{2}(\tau))\pm
F_{n}(\tau)}{2}}, \label{eq:6b}
\end{align}
\end{subequations}
where the function $F_{n}(\tau)$ is
\begin{equation} \label{eq:7}
F_{n}(\tau)=\sqrt{\big(\eta_{\scriptscriptstyle
1}^{2}(\tau)+\eta_{\scriptscriptstyle
2}^{2}(\tau)\big)^{2}+16(n+1)(n+2)\eta_{\scriptscriptstyle
1}^{2}(\tau)\eta_{\scriptscriptstyle 2}^{2}(\tau)}.
\end{equation}
and
\begin{subequations} \label{eq:8}
\begin{align}
\nonumber \left\vert\Psi_{1,2}(\tau)\right\rangle =&
A_{-}(\tau)\vert n,e_{1}e_{2}\rangle+D_{-}(\tau)\vert
n+2,g_{1}g_{2}\rangle \\ &\pm \big(B_{-}(\tau)\vert
n+1,g_{1}e_{2}\rangle-C_{-}(\tau)\vert n+1,e_{1}g_{2}\rangle\big),
\label{eq:8a} \\ \nonumber
\left\vert\Psi_{3,4}(\tau)\right\rangle =& A_{+}(\tau)\vert
n,e_{1}e_{2}\rangle+D_{+}(\tau)\vert n+2,g_{1}g_{2}\rangle \\ &\pm \big(B_{+}(\tau)\vert
n+1,g_{1}e_{2}\rangle-C_{+}(\tau)\vert n+1,e_{1}g_{2}\rangle\big).
\label{eq:8b}
\end{align}
\end{subequations}
The upper sign in Eqs. (\ref{eq:8}) is for the odd numbered
states and the lower one is for the even ones.

\begin{figure}[!t]
  \begin{center}
    \includegraphics[width=\textwidth,height=6.0cm]{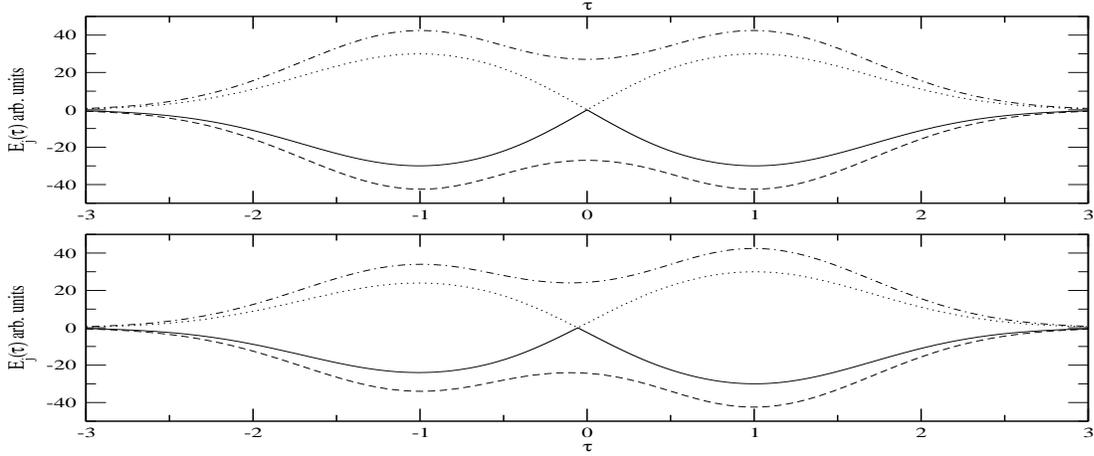}
    \caption{The adiabatic energies. Top: The symmetric limit
      $\epsilon=g_1/g_2=1.0$. Bottom: The asymmetric case for
      $\epsilon=0.8$. The
      energies are $E_1(\tau)$ (solid),
      $E_2(\tau)$ (dot), $E_3(\tau)$ (dashed) and $E_4(\tau)$
      (dot-dashed). The other parameters are: $\delta=1.0$, $n=0$ and
      $\Delta=0$.}
  \label{fig:2}
  \end{center}
\end{figure}

A first important property for the adiabatic states is the temporal degeneracy
for two of them, figure \ref{fig:2}. From Eq. (\ref{eq:6a}) we can see
that for $\eta_1(t)=\eta_2(t)$ we have that $E_1(t)=E_2(t)$. When $g_1=g_2$
then $t=0$ and for $g_1\neq g_2$ then $t$ has a finite value at the degeneracy
which is a function of the ratio $\epsilon=g_2/g_1$.

In the vicinity of this temporal degeneracy the adiabatic approximation will
fail and this is because the adiabatic theorem holds only for
non-degenerate states \cite{Messiah}. One can show that near this
point the system undergoes a pure crossing even for unequal couplings $g_1\neq
g_2$. The reason for this behaviour is the diagonal form that
the Hamiltonian has in the vicinity of the temporal degeneracy, and the fact
that the two adiabatic states, $\left\vert\Psi_1\right\rangle$ and
$\left\vert\Psi_2\right\rangle$, do not couple near the crossing. Thus, if the
system is initially in state
$\left\vert\Psi_1(-\infty)\right\rangle$, $\left(\left\vert\Psi_2(-\infty)
\right\rangle\right)$, then for $t\rightarrow\infty$ the final state will be
$\left\vert\Psi_2(\infty)\right\rangle$, $\left(\left\vert\Psi_1(\infty)
\right\rangle\right)$ with an arbitrary phase factor $\theta_n$, i.e.
\begin{equation} \label{eq:9}
  \left\vert\Psi_{1}(-\infty)\right\rangle\rightarrow e^{-\imu\theta_n}
  \left\vert\Psi_{2}(\infty)\right\rangle,\quad
  \left\vert\Psi_{2}(-\infty)\right\rangle\rightarrow e^{\imu\theta_n}
  \left\vert\Psi_{1}(\infty)\right\rangle,
\end{equation}
and
\begin{equation} \label{eq:10}
  \theta_n = \int_{-\infty}^{0}d\tau'E_{1}(\tau')
  +\int_{0}^{\infty}d\tau'E_{2}(\tau').
\end{equation}

\subsection{Input-output in terms of the bare states for $\Delta=0$} \label{sec:23}
For equal couplings $\epsilon=1$, then $\theta_n=0$.
For this limit, and with the following relations for the coupling functions Eq.
(\ref{eq:2})
\begin{equation} \label{eq:add}
\lim_{\tau\rightarrow\infty}\left(\frac{\eta_{\scriptscriptstyle
1}}{\eta_{\scriptscriptstyle
2}}\right)=0,\qquad\lim_{\tau\rightarrow-\infty}\left(\frac{\eta_{\scriptscriptstyle
2}}{\eta_{\scriptscriptstyle 1}}\right)=0,
\end{equation}
the following input-output table in terms of the bare states (\ref{eq:5}) is
obtained
\begin{subequations} \label{eq:11}
\begin{align}
&\vert n,e_{1}e_{2}\rangle\rightarrow\vert n,e_{1}e_{2}\rangle, \label{eq:11a} \\ \nonumber \\
&\vert n+1,g_{1}e_{2}\rangle\rightarrow-\vert n+1,e_{1}g_{2}\rangle, \label{eq:11b} \\ \nonumber \\
&\vert n+1,e_{1}g_{2}\rangle\rightarrow \cos(\phi_{n})\vert n+1,g_{1}e_{2}\rangle-\imu\sin(\phi_{n})\vert
n+2,g_{1}g_{2}\rangle \label{eq:11c}, \\ \nonumber \\
&\vert n+2,g_{1}g_{2}\rangle\rightarrow \cos(\phi_{n})\vert n+2,g_{1}g_{2}\rangle-\imu\sin(\phi_{n})\vert
n+1,g_{1}e_{2}\rangle, \label{eq:11d}
\end{align}
\end{subequations}
where the angle $\phi_{n}$ reads
\begin{equation} \label{eq:12}
\phi_{n}
=\phi_{4}(\infty)=-\phi_{3}(\infty)=\int_{-\infty}^{\infty}d\tau
E_{4}(\tau).
\end{equation}

Equation (\ref{eq:11b}) describes a complete energy transfer between the two atoms. This will
happen without choosing special values for the system parameters provided we ensure the necessary
conditions for adiabatic evolution. This robust energy transfer is a reminiscent of the STIRAP method
\cite{Bergmann1998,Bergmann1995}.
We also note that Eqs. (\ref{eq:11c}) and (\ref{eq:11d}) describe
a conditional entanglement of the second atom to the field mode. If atom 2 is not excited, then it will be entangled to
the field mode.

An interesting limit for $\phi_{n}$ is the one for $n\gg1$. In this limit one can show that
$\phi_{n}$ has the same dependence with respect to $n$ as the mixing angle in the Jaynes-Cummings model \cite{Lazarou2007}:
$\phi_{n}\approx4g\sigma\sqrt{n\pi}$ for $n\gg1$.
This result suggests that for a large number of photons and with adiabatic
evolution, we will have the same kind of dynamics as in the usual
single atom Jaynes-Cummings
model.
What is different is that
this Jaynes-Cummings rotation is conditional upon the state of the second atom. This could used for
conditional operations in quantum information or for preparing field states with the use of conditional control. Of course
in this limit the field dynamics will be the same as for the Jaynes-Cummings model \cite{Scully}.

\subsection{Applications: Quantum teleportation between cavities} \label{sec:24}

Based on this input-output table applications in the field of Quantum
Information and Quantum teleportation can be realised. We have discussed the
implementation of SWAP and C-NOT gates, generating atomic entanglement and
robust quantum state mapping. The proposed applications are
fairly robust and rely on the control of a single parameter, i.e.\ the mixing angle
$\phi_n$. Furthermore the proposed applications are characterised by
relatively high fidelities up to $99\%$ even for errors of the order of $10\%$
in $\delta$ or $1\%$ in $\sigma$ \cite{Lazarou2007}.

To this end is interesting to consider in detail the realisation of a
teleportation protocol between two different cavities. The setup consists of
three identical cavities placed in a row, with classical EM fields between
them. Assuming that the first cavity is initially in the superposition
\begin{displaymath}
  \alpha\vert0\rangle+\beta\vert1\rangle,
\end{displaymath}
we send a pair of non-excited atoms through the first
cavity, ensuring that $\phi_{-1}=\pi/2$. Then according to Eqs.(\ref{eq:11})
, the state of atom 2 will be
\begin{displaymath}
\alpha\vert g_2\rangle-\imu\beta\vert e_2\rangle.
\end{displaymath}
Using the
EM field after cavity 1 we perform the phase transformation $\vert e_2\rangle
\rightarrow\imu\vert e_2\rangle$. In this way the mode state for cavity 1 is mapped onto atom 2, while atom 1
and cavity 1 are not excited.

The atoms
now cross the second (auxiliary) cavity with arbitrary $\phi_{-1}$.
Subsequently we perform a rotation $-\imu\vert e_1\rangle$ with an EM
field so that we get the state
\begin{displaymath}
\vert0;g_2\rangle\left(\alpha\vert g_1\rangle+\imu\beta\vert e_1\rangle\right).
\end{displaymath}
Finally, the two atoms cross the third cavity with $\phi_{-1}=\pi/2$; the result is to
get both atoms in their ground state and the cavity 3
in the same state as cavity 1 was initially. Thus with this fairly simple
method we can teleport the state of a cavity to another cavity.

For the results up to this point, and for the proposed applications to be valid the
adiabatic approximation must hold. This will be true as long as the
coupling strength is greater than a lower bound defined by the
interaction time and the photon number $n$. For small photon numbers,
$n\sim1$, the coupling strength must be of the order of $10/\sigma$. For
a larger number of excitations in the cavity, this lower bound increases
meaning that the coupling strength or the interaction time must also increase
for the adiabatic approximation to be valid \cite{Lazarou2007}. In addition
the delay time $\Delta t$ must be of the order of the interaction time.
Furthermore, the system is fairly robust with
respect to this parameter, since in our scheme the accurate control of
$\delta$ is not important. Instead a rather simple condition with
$\delta\sim1$, i.~e. $1.0\leq\delta\leq1.25$, must be satisfied as seen in
Ref. \cite{Lazarou2007}.

\section{Effects due to different couplings and finite detuning} \label{sec:3}
\subsection{Different couplings} \label{sec:31}
As already mentioned in section \ref{sec:22} the results for the adiabatic states
and the corresponding energies are general and hold even if
$\epsilon\neq1$. The energy crossing will exist for $g_1\neq g_2$ but the
point where this occurs is no longer at $t=0$.
Because of this the mixing angle $\theta_n$
is no longer zero since the
energies $E_1(\tau)$ and $E_2(\tau)$ are no longer symmetrical; i.e.\
$E_j(-\tau)\neq E_j(\tau),\quad j=1,2$ although $E_1(\tau)=-E_2(\tau)$,
figure \ref{fig:2}. Thus the integral
(\ref{eq:10}) has a finite value. This results in input-output
relations that differ from the ones in Eqs. (\ref{eq:11}).

For example, the state $\vert n;e_1,e_2\rangle$ is equivalent to the following
superposition for $\tau\rightarrow -\infty$
\begin{equation} \label{eq:16}
  -\frac{\left\vert\Psi_1(\tau)\right\rangle+\left\vert\Psi_2(\tau)\right
    \rangle}{\sqrt{2}}.
\end{equation}
For $\tau\rightarrow\infty$ and taking into account the crossing
Eq. (\ref{eq:9}) we have
\begin{equation} \label{eq:17}
  \vert n;e_1,e_2\rangle\rightarrow-\frac{e^{\imu\theta_n}\left\vert\Psi_1(\infty)\right\rangle+e^{-\imu\theta_n}
    \left\vert\Psi_2(\infty)\right\rangle}{\sqrt{2}}.
\end{equation}
Using the limits Eq. (\ref{eq:add}) and Eqs. (\ref{eq:8}) we
have \cite{Lazarou2007}
\begin{equation} \label{eq:18}
  \vert n;e_1,e_2\rangle\rightarrow \cos(\theta_n)\vert
  n;e_1,e_2\rangle-\imu\sin(\theta_n)\vert n+1;e_1,g_2\rangle.
\end{equation}
In a similar way we will find that
\begin{equation} \label{eq:19}
  \vert n+1;g_1,e_2\rangle\rightarrow \cos(\theta_n)\vert
  n+1;e_1,g_2\rangle-\imu\sin(\theta_n)\vert n+1;e_1,e_2\rangle.
\end{equation}

\begin{figure}[!t]
  \begin{center}
    \includegraphics[width=\textwidth,height=6.5cm]{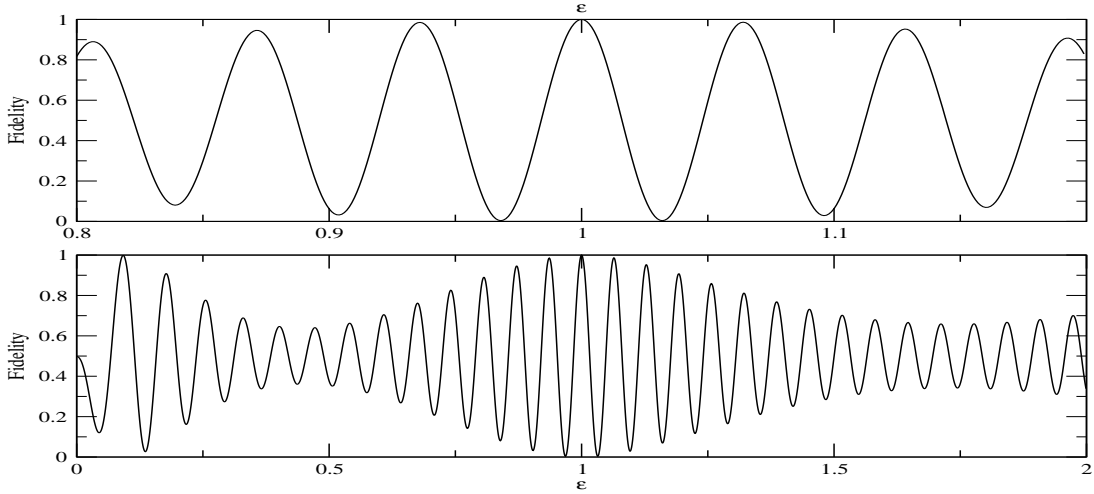}
    \caption{The fidelity for the maximally entangled state
      $\left\vert\Psi_f(0,2m\pi)\right\rangle$ Eq. (\ref{eq:sec31_2}),
      $\left\vert\left\langle\Psi_f(0,2m\pi)\vert\Psi_f(\theta_0,\phi_{-1})\right\rangle\right\vert$
      shown as a function of the asymmetry $\epsilon$.
      The parameters are: $\delta=1$, $\Delta=0$ and
      $g_0\sigma=28.3929$. The coupling $g_0$ was chosen so that for
      $\epsilon=1.0$ the fidelity is unity.
      The upper part of the figure expands the central region in the lower
      part of the figure.
      }
    \label{fig:3}
  \end{center}
\end{figure}

In contrast to Eqs. (\ref{eq:11a}) and (\ref{eq:11b}), we see that there is no
longer robust exchange of energy between the atoms. Furthermore, for
$\epsilon\neq1$ atom 2 will get entangled to the cavity mode without any
condition on its initial state. If it is excited then the entanglement is defined
by the mixing angle $\theta_n$, whereas if it is initially placed in its
ground state the entanglement is defined by the mixing angle $\phi_n$. Both
mixing angles have a dependence with respect to $\delta$, $n$, $\sigma$, $g_0$
and $\epsilon$. For $\theta_n=2\pi$ we recover the same input-output
expressions as Eqs. (\ref{eq:11}).

An interesting feature of the system is that for $\epsilon\neq1$, the robust
state mapping between the two atoms, or the mapping between one of the atoms
and the cavity, which depends on the mixing angle $\phi_{-1}$ and consequently
the teleportation protocol in section \ref{sec:24} remains fairly robust.
The reason for this is that for this application the only states
involved are
\begin{equation} \label{eq:21}
  \vert0;g_1,e_2\rangle,\quad\vert0;e_1,g_2\rangle,\quad\vert1;g_1,g_2\rangle.
\end{equation}
For this subspace the mixing angle $\theta_{n=-1}$ is zero by definition.
Thus the only parameter to be controlled is the mixing angle, as in the
case of identical coupling profiles, $\epsilon=1.0$.

On the other hand, applications, such as the entangling of atoms, become more involved
since an extra control parameter, the angle $\theta_n$, appears in the system
evolution. For example, if the initial state of the system is
\begin{equation} \label{eq:sec31_1}
  \left\vert\Psi_0\right\rangle=\frac{1}{2}\vert0\rangle\left(\vert
  g_1\rangle+\vert e_1\rangle\right)\left(\vert g_2\rangle+\vert
  e_2\rangle\right),
\end{equation}
then the output state will be
\begin{eqnarray}
  \nonumber
  \left\vert\Psi_f(\theta_0,\phi_{-1})\right\rangle&=&\frac{1}{2}
  \left(\vert g_2\rangle\left(\vert
  g_1\rangle-\vert e_1\rangle\right)+\vert e_2\rangle\left(\cos(\theta_0)\vert
  e_1\rangle+\cos(\phi_{-1})\vert
  g_1\rangle\right)\right) \\ & &-\frac{\imu}{2}\left(\sin(\theta_0)
  \vert1;e_1,g_2\rangle+\sin(\phi_{-1})\vert1;g_1,g_2\rangle\right).\label{eq:sec31_2}
\end{eqnarray}
Thus, with a probability that always exceeds $50\%$ the two atoms are
entangled to each other. For example if $\theta_0=2k\pi$ and
$\phi_{-1}=2m\pi$, with $k$ and $m$ integers, the probability is one and the
two atoms are maximally entangled. For different values of the mixing angles
the probability is less than one and the entanglement is not maximal, and
vanishes if $\cos(\theta_0)=0=\cos(\phi_{-1})$.

Because of the complex dependence of both mixing angles with respect to the
asymmetry $\epsilon$, the fidelity of a maximally entangled state is very
sensitive with respect to $\epsilon$. For $\epsilon\sim1$, it oscillates
relatively fast with a decaying amplitude, figure \ref{fig:3}, and revivals
are observed for different values of the asymmetry parameter
$\epsilon$. Despite this, and because the condition
$\cos(\theta_0)=0=\cos(\phi_{-1})$ is only satisfied for certain values of the
asymmetry parameter, the atomic entanglement persists even if the asymmetry
factor is not one.
\subsection{Effects of the finite detuning} \label{sec:32}
\begin{figure}[!t]
  \begin{center}
    \includegraphics[width=\textwidth,height=5.5cm]{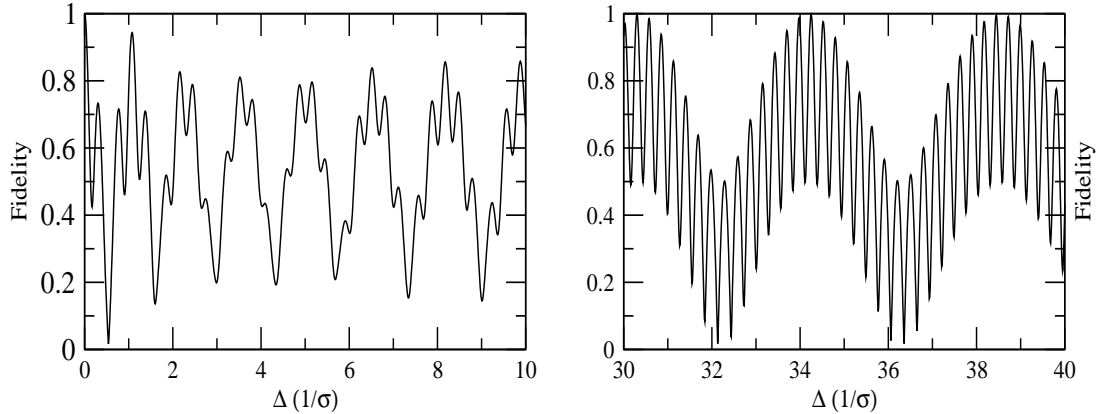}
    \caption{
The fidelity for the maximally entangled state
    $\left\vert\Psi_f^\Delta(2m\pi,0)\right\rangle$ Eq. (\ref{eq:23e}) with respect to the detuning $\Delta$.
    The result was obtained after
    numerical simulations with the Schr\"odinger equation for the
    time-dependent Hamiltonian
    (\ref{eq:4}). The integration interval is
    $-12\sigma\leq t\leq12\sigma$, and the fidelity was calculated for
    $t=12\sigma$. The parameters are: $\delta=1.0$, $\epsilon=1.0$ and
    $g_0\sigma=28.3929$. For this coupling and for
    $\epsilon=1.0$ and $\Delta=0$ the fidelity is unity.}
    \label{fig:4}
  \end{center}
\end{figure}

For finite detuning, one can show that the system has the following dark state
\begin{equation} \label{eq:23}
  \vert D\rangle=\frac{1}{\sqrt{\eta_1^2+\eta_2^2}}\left(\eta_1\vert
  g_1,e_2\rangle-\eta_2\vert e_1,g_2\rangle\right)\vert0\rangle.
\end{equation}
Taking into account Eq. (\ref{eq:add}) we see that the robust energy change
between atoms, Eq. (\ref{eq:11b}), takes place even for a finite detuning. Thus
the robust state mapping between the two atoms can be realised even if the
detuning is not zero. Furthermore for $\Delta\gg g_0$, an effective
Hamiltonian for the states $\vert0;g_1,e_2\rangle$ and $\vert0;e_1,g_2\rangle$
is obtained after adiabatic elimination of the off-resonant states
$\vert0;e_1,e_2\rangle$ and $\vert1;g_1,g_2\rangle$.

This Hamiltonian has two eigenstates (adiabatic states), the dark state $\vert
D\rangle$ Eq. (\ref{eq:23}), and
\begin{equation} \label{eq:23a}
  \left\vert\Psi\right\rangle=\frac{1}{\sqrt{\eta_1^2+\eta_2^2}}\left(\eta_2\vert
  g_1,e_2\rangle+\eta_1\vert e_1,g_2\rangle\right)\vert0\rangle,
\end{equation}
where the corresponding adiabatic energy is
\begin{equation} \label{eq:23b}
  E_{\Psi}(t)=\frac{\eta_1^2+\eta_2^2}{\Delta}.
\end{equation}

Taking into consideration Eq. (\ref{eq:add}) and Eqs. (\ref{eq:23}) and
(\ref{eq:23a}), we get the following input-output relations for the adiabatic
limit
\begin{equation} \label{eq:23c}
  \vert g_1,e_2\rangle\rightarrow-\vert e_1,g_2\rangle,\quad\vert
  e_1,g_2\rangle\rightarrow e^{-\imu\Theta}\vert g_1,e_2\rangle,
\end{equation}
where the angle $\Theta$ reads
\begin{equation} \label{eq:23d}
  \Theta=\frac{2\sigma g_0^2}{\Delta}\int_{-\infty}^{\infty}
  E_{\Psi}(\tau)d\tau=\frac{2\sigma g_0^2(1+\epsilon^2)}{\Delta}\sqrt{\frac{\pi}{2}}.
\end{equation}

Equation (\ref{eq:23c}) represents a SWAP operation, but when it is combined with
the initial state $\left\vert\Psi_0\right\rangle$, Eq. (\ref{eq:sec31_1}), it
gives the following output state
\begin{equation} \label{eq:23e}
  \left\vert\Psi_f^\Delta(\Theta,\Phi)\right\rangle=\frac{1}{2}\vert0\rangle\left(\vert
  g_2\rangle\left(\vert g_1\rangle-\vert e_1\rangle\right)+\vert
  e_2\rangle\left(e^{-\imu\Theta}\vert g_1\rangle+e^{-\imu\Phi}\vert
  e_1\rangle\right)\right).
\end{equation}
The phase factor $e^{-\imu\Phi}$ represents the phase acquired by the state
$\vert0;e_1,e_2\rangle$ during the system evolution.
This is an entangled state of the two atoms and becomes a maximally entangled
state if the two phases are both $e^{-\imu\Theta}=e^{-\imu\Phi}=\pm 1$ since the
fidelity for the entangled state (\ref{eq:23e}) is
\begin{equation} \label{eq:23f}
  F=\frac{1}{4}\left\vert2+e^{-\imu\Theta}+e^{-\imu\Phi}\right\vert,
\end{equation}
and it has a maximum, $F=1$, for $\Theta=2m\pi$ and $\Phi=2k\pi$ where
$m,k=\pm1,\pm2\dots$.

\begin{figure}[!t]
  \begin{center}
    \includegraphics[width=\textwidth,height=5.5cm]{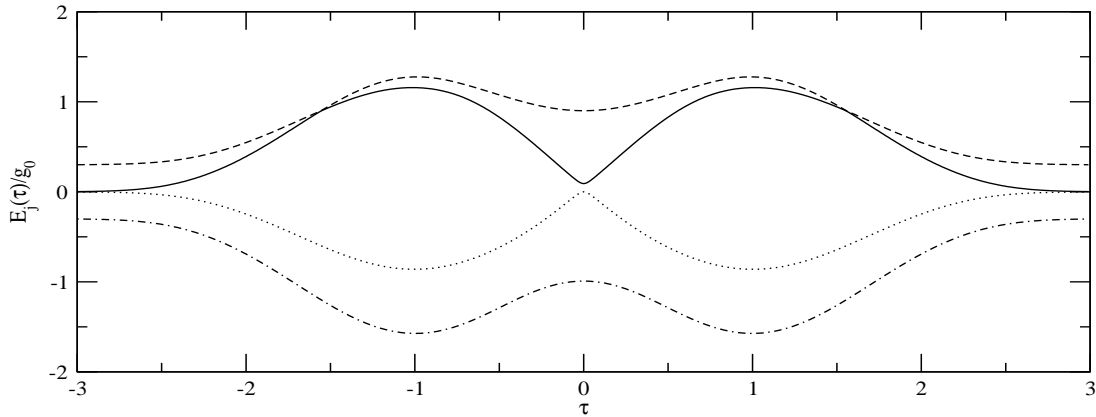}
    \caption{The adiabatic energies for $\Delta=15/\sigma$:
      $E_1(\tau)$ (dot), $E_2(\tau)$ (solid), $E_3(\tau)$ (dot-dashed) and
      $E_4(\tau)$ (dashed). The parameters are: $\delta=1.0$, $n=0$ and
      $\epsilon=1.0$. Notice
      the three avoided crossings for $\tau=0$ and the two symmetric with
      respect to $\tau=0$.}
    \label{fig:5}
  \end{center}
\end{figure}
In figure \ref{fig:4}, the fidelity is plotted with respect to the detuning
$\Delta$ as calculated after numerically integrating the Schr\"odinger equation
for the Hamiltonian (\ref{eq:4}). We see that for a small detuning the fidelity
decays fast, below $50\%$, and then oscillates with a non harmonic profile,
and an average value below unity, figure \ref{fig:4} (left). This is due
to the fact that either both atoms, or one of them, entangles to the
cavity, and thus we don't have pure atomic entanglement.
As the detuning increases the fidelity continues oscillating, but the
average fidelity varies periodically between zero and one; figure
\ref{fig:4} (right). Subsequent investigation has shown that
these variations in the average fidelity are due to the interference
between the two phase terms involving
$e^{-\imu\Phi}$ and $e^{-\imu\Theta}$.

\begin{figure}[!t]
  \begin{center}
    \includegraphics[width=\textwidth,height=5.5cm]{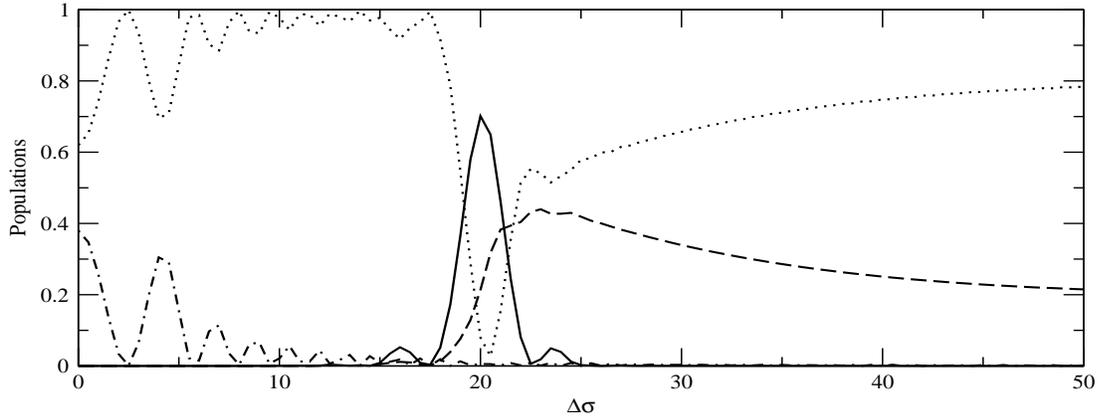}
    \caption{The final populations for the states $\vert0;e_1,e_2\rangle$
      (solid), $\vert1;g_1,e_2\rangle$ (dot), $\vert2;g_1,g_2\rangle$
      (dot-dashed) and $\vert1;e_1,g_2\rangle$ (dashed). This latter state was
      chosen to be the initial state of the system. The parameters are:
      $\delta=1.0$, $n=0$, $g_1\sigma=50$, $\epsilon=1.0$ and $-6\sigma\leq\
      t\leq6\sigma$. The populations were calculated at $t=6\sigma$. Notice
      the evidence for tri-partite entanglement when $\Delta\sigma\sim20$.}
    \label{fig:6}
  \end{center}
\end{figure}

In general, the off-resonance case is characterised by three regimes. The
first regime is that for small detuning, $\Delta/\sigma<20$, where the
system qualitatively behaves in a similar way
to the resonant case. More specifically the second atom entangles to the
cavity mode, as is evidenced by the populations seen in
figure \ref{fig:6}. The entanglement depends on $\Delta$ and
equations (\ref{eq:11}) are no longer valid.

For $\Delta\sim20/\sigma$, the adiabatic spectrum has three avoided
crossings, figure \ref{fig:5}. Because of this the two atoms entangle
to the cavity mode forming the state
\begin{equation} \label{eq:22}
  c_1\vert n;e_1,e_2\rangle+c_2\vert n+1;e_1,g_2\rangle+c_3\vert
  n+1;g_1,e_2\rangle.
\end{equation}
Each of the three adiabatic states which are involved in the avoided crossings, figure
\ref{fig:5}, map, at
$\tau\rightarrow\pm\infty$, with one of the three states appearing in
Eq. (\ref{eq:22}). Because of the small gap in the vicinity of the avoided
crossings, the three adiabatic states are coupled to each other and as a result
the system, which starts in one of the three states, $\vert n;e_1,e_2\rangle$,
$\vert n+1;e_1,g_1\rangle$ or $\vert n+1;g_1,e_2\rangle$, ends up in an entangled
state similar to (\ref{eq:22}), figure \ref{fig:6}.

In the limit of large detuning, $\Delta\geq g_0$, as already discussed, the
cavity is not excited and does not entangle to the atoms, as is shown by
the populations seen figure \ref{fig:6}.
The two atoms can interact with each other via virtual
excitations of the cavity field, and get entangled. Similar
results were previously obtained for $\Delta\gg g_0$ with use of a time dependent
Fr\"olich transformation \cite{Yong2007}.
\clearpage

\section{Effects due to cavity losses} \label{sec:4}

The results from the previous sections were derived without taking into
account the cavity decoherence due to photon losses.
In order to understand the importance of decoherence, we solved the master
equation for the density matrix $\rho(t)$ of the entire atom-cavity system,
\begin{equation} \label{eq:24}
  \frac{d\rho}{dt}=-\imu\left[H_{I}(t),\rho\right]+\mathcal{L}(\rho).
\end{equation}
The term $\mathcal{L}(\rho)$ describes the damping of the field mode with a rate
$\gamma$ into an empty thermal reservoir at zero temperature \cite{Scully}
\begin{equation}
  \mathcal{L}(\rho)=-\frac{\gamma}{2}\left(a^\dagger a\rho+\rho a^\dagger
  a-2a\rho a^\dagger\right).
\end{equation}

\begin{figure}[!t]
  \begin{center}
    \includegraphics[width=\textwidth,height=5.5cm]{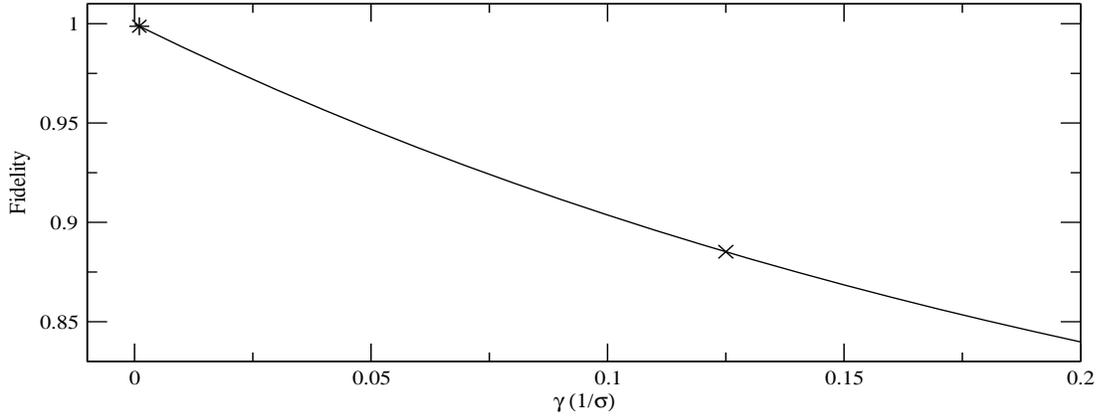}
    \caption{The fidelity for the maximally entangled state
    $\left\vert\Psi_f(0,2\pi)\right\rangle$ Eq. (\ref{eq:sec31_2}) (solid
    line) with respect to the decay rate $\gamma$. The curve was found by
    integrating Eq. (\ref{eq:24})
    for many different decay rates with $\Delta\gamma=0.01/\sigma$. The special
    points marked $(\ast)$ and $(\times)$ are discussed in detail in the
    text. The parameters are: $\delta=1.0$, $\epsilon=1.0$,
    $g_0\sigma=18.9286$, $\Delta=0$ and $-12\sigma\leq t\leq12\sigma$. The
    fidelity was calculated for $t=12\sigma$. For this choise of coupling the
    fidelity is equal to one for $\gamma=0$.}
    \label{fig:7}
  \end{center}
\end{figure}

The main result of the simulations with Eq. (\ref{eq:24}), is that,
as expected, the
predictions of the previous sections will hold as long as the cavity losses
are substantially suppressed. For example, in figure \ref{fig:7} the fidelity
for the maximally entangled state is plotted with respect to $\gamma$. As long as
the decay rate $\gamma$ is much smaller than $1/\sigma$, then the fidelity
remains large, of the order of unity. For a micromaser cavity with $Q\sim10^7$,
a photon lifetime $160\mu s$ and $\sigma\sim20\mu s$ \cite{Brune1996},
the decay rate is approximately $\gamma\sigma=0.125$. For such a cavity the
fidelity is just less than $0.9$, figure \ref{fig:7}.
On the other hand, for a micromaser cavity with $Q\sim10^{10}$, a photon
lifetime is $0.1s$, and the interaction time $\sigma$  of the order of $100\mu s$
\cite{Varcoe2004}, the decay rate is approximately $\gamma=10^{-3}/\sigma$. For
this decay rate the fidelity is $F\approx1-10^{-4}$. Both cases
are shown in figure \ref{fig:7}: $\gamma\sigma=10^{-3}$ marked with an
asterisk and $\gamma\sigma=0.125$ with a cross. Thus, the system dynamics
can be
well described in terms of the ideal model in the absence of
decoherence in practical cases, such as the
high $Q$ cavity example with $Q\sim10^{10}$.

\section{Conclusion} \label{sec:5}

In this paper, we examine a system of two atoms interacting with a single mode
cavity. Assuming that the atoms enter the cavity at different times, and
follow different trajectories inside the cavity, we utilise a cavity-atom
interaction with sequential time dependent coupling profiles.
We have studied the importance of asymmetries between the
atomic coupling profiles. In general the behaviour of the unbalanced system is similar to
the ideal case when the coupling profiles are identical in shape. The atom that enters
the cavity second is entangled to the cavity mode, whereas the first atom is
not. This is due to the existence of an energy crossing in the
adiabatic spectrum. Furthermore, the entanglement is defined by the initial
state of the second atom, and the degree of entanglement is a function of
two mixing angles. Both mixing angles are functions of the coupling strength,
the interaction time and the asymmetry factor.

The proposed teleportation protocol remains \emph{fairly} robust since the
system control is a function of only one mixing angle. This is due to the fact
that the states involved in the protocol belong to a subspace of the general
Hilbert space with a single excitation. Within this subspace, one of the
mixing angles is by definition zero. On the other hand, the generation of
a maximally entangled state is rather sensitive to variations of the asymmetry
factor, with an intense oscillatory fidelity. For an asymmetry factor
$\epsilon\approx0.99$ the fidelity drops to $0.8$, where as for
$\epsilon\approx0.96$ the fidelity is zero.

For off-resonance interactions, the system is characterised by three distinct
dynamic regimes. For small detuning, the system qualitatively behaves in a
similar way as in the resonant limit. The second atom entangles to the cavity
where the entanglement is a function of the detuning and the remaining system
parameters. For moderate detunings, both atoms entangle to the cavity this being
due to three avoided crossings in the adiabatic spectrum. For a detuning
larger than the coupling, the cavity decouples from the system evolution and
atoms interact with each other via virtual excitations of the cavity. The
system evolves inside a two dimensional phase space, and the fidelity for a
maximally entangled state has an extreme behaviour with respect to one of the
phase parameters. On the other hand even with a finite detuning the robust
state mapping between atoms is valid.

A potential experimental realization of the current model requires a high
$Q$ micromaser in order to suppress the cavity losses. A cavity with small
quality factor, $Q\sim10^7$ is in practice substantially affected by
decoherence effects reducing the fidelity of the proposed applications by a
factor greater than $10\%$. On the other hand, for a cavity with $Q$ of the
order of $10^{10}$, the decoherence is found to have negligible effects on
the system evolution. For example, with such a cavity the fidelity of a
maximally entangled state reduces only by $0.2\%$.
\begin{acknowledgement}
BMG acknowledges support from the Leverhulme Trust.
\end{acknowledgement}
\bibliographystyle{epj}
\bibliography{Lazarou2008.bbl}

\begin{thebibliography}{21}

\bibitem{Nielsen}
M.A. Nielsen, I.L. Chuang, \emph{Quantum computation and quantum information}
  (Cambridge University Press, Cambridge, 2000)

\bibitem{Zheng2005}
S.B. Zheng, Phys. Rev. A \textbf{71}(6), 062335 (2005)

\bibitem{Zheng2000}
S.B. Zheng, G.C. Guo, Phys. Rev. Lett. \textbf{85}(11), 2392 (2000)

\bibitem{Jane2002}
E.~Jan\'e, M.B. Plenio, D.~Jonathan, Phys. Rev. A \textbf{65}(5), 050302 (2002)

\bibitem{You2003b}
L.~You, X.X. Yi, X.H. Su, Phys. Rev. A \textbf{67}(3), 032308 (2003)

\bibitem{Marr2003}
C.~Marr, A.~Beige, G.~Rempe, Phys. Rev. A \textbf{68}(3), 033817 (2003)

\bibitem{Yong2007}
L.~Yong, C.~Bruder, C.P. Sun, Phys. Rev. A \textbf{75}(3), 032302 (2007)

\bibitem{Plenio1999}
M.B. Plenio, S.F. Huelga, A.~Beige, P.L. Knight, Phys. Rev. A \textbf{59}(3),
  2468 (1999)

\bibitem{Beige2000a}
A.~Beige, S.~Bose, D.~Braun, S.~Huelga, P.~Knight, M.~Plenio, V.~Vedral,
  Journal of Modern Optics \textbf{47}, 2583 (20 November 2000)

\bibitem{Beige2000b}
A.~Beige, D.~Braun, B.~Tregenna, P.L. Knight, Phys. Rev. Lett. \textbf{85}(8),
  1762 (2000)

\bibitem{Shorencen2003}
A.S. S\o{}rensen, K.~M\o{}lmer, Phys. Rev. Lett. \textbf{90}(12), 127903 (2003)

\bibitem{Chen2003}
T.W. Chen, C.K. Law, P.T. Leung, Phys. Rev. A \textbf{68}(5), 052312 (2003)

\bibitem{Duan2003}
L.M. Duan, H.J. Kimble, Phys. Rev. Lett. \textbf{90}(25), 253601 (2003)

\bibitem{Lazarou2007}
C.~Lazarou, B.~Garraway, Adiabatic entanglement in two-atom cavity QED.
  Submitted 

\bibitem{Messiah}
A.~Messiah, \emph{Quantum Mechanics} (Dover Publications, New York, 1999)

\bibitem{Mahmood1987}
S.~Mahmood, M.S. Zubairy, Phys. Rev. A \textbf{35}(1), 425 (1987)

\bibitem{Bergmann1998}
K.~Bergmann, H.~Theuer, B.W. Shore, Rev. Mod. Phys. \textbf{70}(3), 1003 (1998)

\bibitem{Bergmann1995}
K.~Bergmann, B.W. Shore, in \emph{Molecular Dynamics and Spectroscopy by
  Stimulated Emission Pumping}, edited by H.L. Dai, R.W. Field (World
  Scientific, Singapore, 1995), chap.~9, pp. 315--73

\bibitem{Scully}
See for example M.O. Scully, M.S. Zubairy, \emph{Quantum Optics} (Cambridge University Press,
  Cambridge, 2002)

\bibitem{Brune1996}
M.~Brune, E.~Hagley, J.~Dreyer, X.~Ma\^itre, A.~Maali, C.~Wunderlich, J.M.
  Raimond, S.~Haroche, Phys. Rev. Lett. \textbf{77}(24), 4887 (1996)

\bibitem{Varcoe2004}
B.T.H Varcoe, H.~Walther, New Journal of Physics \textbf{6}, 97
  (2004)

\end{thebibliography}
\end{document}